# The atomic hypothesis

*S. K. Bose*

We propose to review the origin and development of the atomic hypothesis. By this is meant the idea that the phenomenal world around us is made up, ultimately, of minute constituent entities that cannot be further subdivided. These are the atoms. The hypothesis of atoms arose in two classical civilizations: Greek and Indian. The contribution of the Greeks is well known while that of Indians is hardly known. Accordingly, we will devote greater effort to explore the Indian contribution.

The first step in reducing the complexity of the phenomenal world is to enquire if the apparently endless variety of objects, both living and non-living, could be explained in terms of a few basic constituent elements. At least three ancient civilizations grappled with this question and came up with strikingly similar answers. Let us briefly recall these developments.

In the Western intellectual tradition, Thales is acknowledged to be the father of philosophy[1]. He lived in Miletus, a city in Asia Minor in present-day Turkey during the 6th century BCE. He is reputed to have predicted an eclipse that astronomers estimate must have taken place in 585 BCE. Thales surmised that the fundamental element constituting the world was water. A second member of the Milesian school, Anaximander (610–546 BCE), hypothesized that the primary element was all-pervading, limitless, boundariless and eternal, which he called *apeiron*. Somewhat later, Anaximenes, also from the Milesian school, asserted that air was the primary element from which everything else is derived. Another Greek philosopher deserving mention is Xenophanes, who declared that everything is made up of earth and water[1]. The next important Greek philosopher was Heraclitus[1], according to whom fire was the primordial element that gave rise to everything else (note 1). Heraclitus flourished around 500 BCE (note 2) and was from Ephesus in Asia Minor. Finally, Empedocles advanced the idea[1], in around 440 BCE, that the primary elements are four in number; these being water, earth, air and fire. This was accepted by all later thinkers.

Aristotle did introduce a fifth element; however, it did not take part in the constitution of objects around us (see below).

Philosophical speculation concerning the nature of things begins, in the Indian tradition, with the Upanisads. These were composed towards the end of the Vedic period, that is, around 600 BCE. In the Brhad-aranyaka Upanisad, perhaps the earliest of the Upanisads, it is stated that water is the primal element[2]. In fact, it goes on to state that out of water is born earth and then fire. In the Chandogya Upanisad space (*akasa*) is declared to be the primal element. Space is said to be the origin, support and the final end of all[2]. These explorations led to the concept of five primary elements (pancabhuta), as stated in the Taittiriya upanisad[2]. These elements are space (*akasa*), water (*ap*), fire (*agni*), air (*vayu*) and earth (*prithvi*). Each of these elements has its own characteristic property. Thus, space has the quality of sound (*sabda*), water that of taste (*rasa*), fire that of colour (*rupa*), air of touch (*sparsa*) and earth of odour (*gandha*). These qualities are associated with the fivefold scheme of sense organs, namely those of hearing, taste, sight, touch and smell respectively. It would thus appear that the concept of pancabhuta is essentially of empirical origin and not the result of pure speculation. In the Svetasvatara Upanisad, the acceptance of the concept of five elements is implied[2].

The Chinese analogue of the Greek and Indian ideas, sketched above, is the theory of five elements. This was developed sometime during 350–270 BCE, when Tsou Yen systematized the ideas that had been in vogue for more than a century[3]. The five elements are water (note 3), fire, wood, metal and earth. Each of these elements has an associated process that occurs in nature. Thus, water has the process that we describe as soaking and descending. Fire has the process that we describe as blazing and uprising. Wood has the quality that permits straight edges or curved surfaces. Metal can follow the shape of a mould and then become hard. Earth allows sowing, growth and reaping. The five-elements concepts led to the idea of five colours, five flavours, five musical modes and other fivefold.

The four basic elements of Empedocles are the same as the material elements of the Indian pancabhuta. The latter has also space. The use of space may have an element of prescience in that it might have led to the concept of 'atoms and void' as constituting apparently solid matter. Also, the upanisadic concept of space (akasa) bears striking resemblance with Anaximander's apeiron. Further, akasa should not be confused with the concept of the Brahman (or Chinese concept of the Tao). The latter does not constitute anything, rather it is a principle that is immanent in and behind all things. At this stage it is useful to take note of a somewhat later development. Aristotle (384–322 BCE) introduced a fifth element, but only as a constituent of heavenly bodies. According to his scheme, the Earth is at the centre of the universe, above it is the Moon and beyond the moon is are the Sun and other heavenly bodies. The sublunar part of creation is made out of the four elements of Empedocles. The remaining heavenly bodies are made only out of the fifth element. The Chinese theory of five elements involves only material things; three of these, water, fire and earth, are the same as in Indian and Greek schemes; while the remaining two, namely, wood and metal, are distinct. This situation is probably connected with the very high level of technological development achieved in China.

The fact that distinct civilizations reached similar conclusions is quite remarkable. It is an intriguing question as to what it implies. If nothing else, it suggests a certain universality of the human intellect in its endeavour to understand the workings of nature.

One cannot fail to notice a feature of the Indian scene; we cannot always associate a name with an idea. A certain amount of reticence in claiming credit for oneself seems to have been always present in the Indian tradition. Thus, from an analysis of the linguistic and literary styles, scholars have concluded that the *Bhagavad-Gita* did not form part of the original epic *Mahab-* *harata*, but is a later insertion[4]. But who composed the *Bhagavad-Gita*? To cite another example, what is the name of the artist who created the exquisite kangra miniatures illustrating Jayadeva's *Gita-Govinda*? (note 4)[5]. For that matter, who is the painter responsible for the portrait of the musician Himmat Khan Kalawant playing the rudra-veena (note 5)[6] – unquestionably a masterpiece in the realm of art?

The idea that an apparently continuous mass distribution consists really of discrete atoms and space arose as follows. Let us first assume that matter cannot be destroyed or created (in the sense that was enshrined two millennia later in the Newtonian mass conservation law). Now it is something of elementary experience that a solid piece of material can be broken into two pieces. This implies that there must have been space in the material along which the breakage proceeds. Another way of stating the same is this. If I insert a knife in a slab of butter, all that is happening is that, since no amount of matter can be destroyed, the knife has gone into the space that already existed in the slab. In other words, if the material was truly continuous, it would not permit a foreign body to be inserted into it; that is, it would offer infinite resistance to such an effort. If we imagine this process of breakage taking place repeatedly, we then come finally to a stage when further subdivision of our sample is no longer possible. We are left with entities that are indivisible and indestructible. These are the atoms.

The earliest explicit articulation of the atomic doctrine is from Pakuda Kacchayana, a contemporary of the Buddha (566–486 BCE)[7]. He is briefly mentioned in the canonical Buddhist text *Digha Nikaya* (note 6)[8], along with five other heterodox (those that reject the authority of the Vedas) preachers, who were leaders of their respective sects of recluses and laymen. This occurs in the course of a conversation between the King Ajatsatru of Magadh and the Buddha, wherein the king recalls his prior discussions with the six preachers. Here is a summary of Pakuda's teaching: 'The following seven things are neither made nor commanded to be made, neither created nor caused to be created, they are barren (so that nothing is made out of them), steadfast as a mountain peak, as a pillar firmly fixed. They move not, neither do they vary, they trench not one upon another, nor avail aught as to ease or pain or both. And what are the seven? The four elements – earth, water, fire and air – and ease, and pain, and the soul as the seventh. So there is neither slayer nor causer of slaying, hearer or speaker, knower or explainer. When one with a sharp sword cleaves a head in twain, no one thereby deprives any one of life, a sword has only penetrated into the interval between seven elementary substances' (note 7)[9].

To get a feel for the intellectual ferment of the time when the Buddha walked on this earth, we recall the five remaining thinkers mentioned in the *Digha Nikaya*. These are: Purana Kassapa, Ajita Kesakambali, Makkhali Gosala (also Maskarin Gosala), Nigantha Nataputta and Sanjaya Belatthiputta. Other than Ajita, all accepted the prevailing Indian concepts of transmigration, rebirth and the ultimate aim of release from the cycle of rebirth; and the associated law of karma. However, they rejected the need of any divine intervention for implementation of this law; all of them being atheists to a degree. Purana taught the doctrine that virtuous conduct (or lack thereof) had no effect on a man's karma. Makkhali was the leader of a clan – Ajivikas – that survived for almost two millennia before becoming finally extinct. Ajivika teaching has many things in common with that of the Jains, as we will shortly discuss. Ajita was an uncompromising materialist, probably the first in the Indian tradition, who rejected the idea of transmigration, rebirth and suchlike, and to whom a man's death was the end of it all. Sanjaya was the skeptic – the doubting Thomas of his age – who denied the possibility of ever knowing anything for certain. Nigantha Nataputta is more famous under the name Vardhaman Mahavira (note 8) – the last in a line of 24 teachers (*tirthankaras*) in the Jain tradition. The Jain tradition survives and is a vital presence in the spiritual landscape of India.

The heterodox doctrines of Ajivika, Jain and followers of the Buddha all accept the atomic hypothesis. For the former two traditions, it is difficult to decide exactly when the idea was propounded. Mahavira was a contemporary of the Buddha, but he is not considered to be the original teacher of the doctrine. He is, rather, a resuscitator who is believed to have revived a much earlier tradition. Gosala is supposed to have joined a band of recluses – the Ajivikas, and then finally taken over as their leader. In any case, the Jain cosmography is as follows: the universe consists of the Jiva and Ajiva (not Jiva). Jiva is the aggregate of all the life-monads (note 9); it is a substance of adjustable magnitude that can be housed, at a given time, in various objects such as plants, insects, animals and finally humans (note 10). Ajiva consists of space (three types, akasa, that which allows objects to be contained, and two others associated with motion and with rest), time and matter (*pudgala*). Pudgala appears in manifold forms, the ultimate stage of it is atomic. The atoms are all of only one type. The distinction between various material objects that we see around us is due to the different ways in which the atoms combine to make these objects. Finally, the individual atom cannot be detected with our sense organs; only in aggregates to they become amenable to detection. All these atoms and their aggregates are supposed to harvour life-monads so that the Jain universe is one that is throbbing with life[10].

The Ajivikas also take the atomic doctrine for granted, without fuss, without much ado, just like the Jains. This probably implies, as Zimmer[11] has argued so persuasively, that the idea itself has roots that go much further back, possibly to the pre-Vedic, pre-Aryan times. The main Ajivika concern is, however, with evolution. A life-monad starts out its journey inside an atom and then progressively evolves via transmigration and reincarnation into vegetation and plant life, into stages of insect and animal lives and finally the stages in the human sphere. All the life-monads of the universe are laboriously treading along this path. After exactly eighty-four hundred thousand existences comes the moment of liberation, automatically. The natural course of this evolution, controlled by an all-pervasive cosmic principle called *niyati*, cannot be influenced in any way by human intervention. Good deeds cannot hasten nor bad deeds impede the process of reaching liberation. It is here that the Ajivika doctrine differs from the Jain. According to the latter (as well as to the Hindu or the Buddhist doctrines), good deeds on the part of the individual do bring in rewards concerning the escape from the cycle of rebirth. Finally, it is interesting to note that the concept of eighty-four hundred thousand existences (Chaurasi lakh Janam) is present even today, e.g. in the sacred book, Guru Granth, of the Sikhs, it is mentioned many times.

Let us now come back to the *terra firma* of Greek philosophy[1]. Leucippus of Miletus (440 BCE) is generally credited with the first formulation of the atomic hypothesis. However, we know very little of this somewhat shadowy figure; so much so that a later atomist, Epicurus, even doubted his very existence. In any event, the atomic hypothesis was further developed by Democritus (around 420 BCE), who erected a fairly elaborate scheme of nature on this basis. We know of Democritus much better. He was a contemporary of Protagoras – the father of Sophism. Incidentally, the word sophist did not carry the somewhat unsavoury connotation that it has today. Originally it simply meant a philosopher or possibly a teacher of philosophy.

In the scheme of Democritus everything in the world is made up of atoms and the void, with the void filling up the intervals between the atoms. The atoms are indivisible and cannot be destroyed or created. Atoms come in distinct types, infinite in number, their distinction arising out of differences in size and shape. Aristotle also attributes to the atomists the concept of heat associated with atoms; the spherical atoms that make up fire being the hottest. The idea of infinite variety of atoms is, quite clearly, the least attractive part of the scheme; since it abandons the eternal scientific quest of trying to explain the apparently limitless multiplicity in nature in terms of a few conceptual categories. Democritus' atoms are always in motion, somewhat like the picture presented in the modern kinetic theory of gases (at finite temperature). Sometimes two atoms will impinge upon each other, and if their shapes are compatible, will interlock to give rise to a more complex structure. Repeated occurrences of this sort are believed to have given rise to bulk matter. Finally, a word about Democritus theory of the soul. The soul consists of the finest and most mobile atoms and permeates the body wherein it is housed. It is the soul that produces the phenomenon of life. This concept is strikingly similar to the Jain concept of the Jiva that we discussed above. However, we recall that Jain Jiva is not atomic, only Ajiva is.

The next Greek atomist is Epicurus. According to him, the central purpose of human existence is to have pleasure. And pleasure arises, principally, from knowing how the world works. In other words, from knowledge of philosophy and the sciences, and the pursuit after such knowledge (note 11). Leading a life of pleasure will result in a state of tranquillity (*ataraxia*) and freedom from fear of pain and death. In later times Epicureanism came to be associated with unbridled pursuit of the pleasures of the flesh. This is almost certainly the result of smear hurled by the followers of rival, possibly, non-materialistic doctrines. It is interesting to note that a similar development took place in India as well. The materialistic tradition started by Ajita Kesakambali was nurtured by a succession of teachers who came to be known as the Lokayata school. The last great figure of this tradition is Carvaka, who was successfully turned into an object of ridicule – one whose doctrine supposedly is 'of a contented stomach and no thinking' – by his non-materialistic, theistic rivals. Epicurus apparently authored 37 books, of which only fragments survive. Our knowledge of his teachings is almost entirely based on the work of the Roman poet Titus Lucretius Carus, who expounded the Epicurean doctrine in six books of Latin verse entitled *De Rerum Natura*[12]. Here, the atomic hypothesis of Democritus is accepted but not his concept of atomic motion. Instead, the picture here is one of atoms perpetually falling through an infinite void, along parallel straight-line paths. On this is superimposed an additional random motion: occasionally an atom will experience a 'swerve' which will cause it to deviate from its original path and hence collide with another atom. If the shapes of the two atoms are compatible they would interlock, thus leading to the formation of a more complex (diatomic) structure. This process, repeated a sufficient number of times, is believed to have led to the formation of the world.

Although the picture of 'atoms falling through void' is more poetic than scientific and no explanation is provided as to the cause of the 'swerve' (ref. 12), the picture may still have a germ of truth in it. Consider a self-gravitating gas of particles 'falling' towards the centre of mass of the gas along rectilinear (but not parallel!) paths. Consider a superimposed motion, a swerve, due to thermal agitation. Now we have a model for the formation of a massive body. Having thus explained the birth of the world, Lucretius then goes on to advance the idea that life also owes its origin to the interaction of atoms, a staggeringly original and bold hypothesis for that time. At any rate, there is no 'spark of life' of divine origin. This was followed by a remarkably realistic portrayal of the life of early humans, as also the scientific explanation of the phenomena of thunderbolts and volcanic activity. All these to free the human mind from the fear of whimsical actions of capricious gods. But this was not to be. The rationalist world-view of atomists would soon be replaced by Platonic phantasies and then by Old Testament fairy tales. Europe had to wait for a long time before rediscovering her Greek heritage. A somewhat similar development, concerning the decline of the rationalist *Weltanschauung*, took place in India as well. As we have described in the foregoing, the idea of a life-monad undergoing successive stages of evolution was advanced by the Jain and Ajivika thinkers. The idea was also accepted by the Buddhists and the Hindus (note 13), although there were differences amongst them regarding matters of detail. The rationalist viewpoint got eventually superceded by the creeping advance of orthodox, theistic dogmas. As a result, Indian thinking failed to rise from poetic indulgence to the level of critical enquiry that could lead to the development of science. Indeed, India continues to be under the thralldom of irrational superstitions.

Indian atomism reached its apogee in the *Vaisesika sutra* of Kannada[9,10,14]. This work is classified among the six orthodox (one that accepts the authority of the Vedas; or at least pays lip service to it, as the present work certainly does) schools of Indian philosophy. There is some uncertainty regarding the time of composition of this work. Most scholars agree that *Vaisesika* came after Buddha, whereas Roy (note 14)[15] has argued that it precedes and, in fact, influenced the Buddha. In any case, Vaisesika is mentioned (note 15)[16] in the *Arthasastra* – a treatise on theories of statecraft authored around 300 BCE by *Kautilya*, who was the Prime Minister of Chandragupta, the first emperor of the Mauryan dynasty at Pataliputra. Thus, we can safely ascribe the composition of *Vaisesika sutra* to the interval between 486 (Buddha's death) and 300 BCE.

The concept of the atom is arrived at via the process of inference in *Vaisesika*.

# HISTORICAL NOTES

It is a matter of common experience that a piece of gross matter can be divided and subdivided, with each subsequent piece maintaining the properties of the original sample. Can the process of division be carried out *ad infinitum*? If the answer to this question is yes, then it would imply that every object is made up of an infinite number of constituent units and there will be no way of distinguishing between a mountain and a molehill. Thus the process of continued division of an object must stop after a finite number of steps and the last constituent in this process is termed the atom (*paramanu*). Being further indivisible, they cannot be destroyed (or created). Thus an atom occupies a non-zero spatial volume. This then would account for the difference between a mountain and a molehill in terms of the difference in the number of their constituent atoms. It is also implicitly accepted that atoms are spherical objects; since nothing is said of their shape. Although of finite extension, an atom is too small to be visible; that is, to be amenable to direct observation through our sense organs. On the other hand, suitable aggregates of atoms are obviously visible. When does an aggregate begin to be visible? According to *Vaisesika*, there are exactly four types of atoms: those of water, fire, air and earth (pancabhuta minus space/aether). Two atoms of the same type can combine (conjoin) to form a dyad. Thus, there are four types of dyads. Three dyads of the same type form a triad. A triad is sufficiently large in size to be visible, 'as motes in a sunbeam' (this striking imagery appears also in Lucretius). Triads of the same type as well as those belonging to different types can form larger aggregates. It is out of the latter that the whole visible universe, including bodies of living objects, is made.

The picture of an atom in Vaisesika is thus one of a little hard ball. In that case how does a water-atom differ from a fire-atom? Now, bulk matter of different elements (bhuta) have distinct qualifying properties: water, fire, air and earth respectively, those of taste, colour, touch and odour. These qualities must ultimately arise, via inherence, from those of their respective atoms. Thus there must be a distinctive characteristic (*visesa guna*) in each atom that differentiates atoms of one variety from those of another (note 15). A point to note here is that in the Jain atomism, all the four *gunas* are attributed to each atom; since in this tradition there is only one type of atom. A striking fact about Vaisesika doctrine is that not only are atoms belonging to distinct classes different from each other; even different atoms in one and the same class are asserted to be so. Thus an atom is believed to be endowed with a kind of individuality. Actually, the idea is not as esoteric as it appears. To take a modern analogy, it is true in the classical Boltzman statistics, where identical particles are distinguishable. The property is no longer valid in quantum statistics – Bose–Einstein or Fermi–Diract.

Vaisesika does not require the atoms to be necessarily in perpetual motion. This is in contrast to the position of Epicurus–Lucretius. Recall that motion was essential in the latter scheme; since it is motion that led to collisions that, in turn, led to the formation, via interlocking, of more complex structures. This may very well be due to Greek reluctance to accept the idea of action-at-distance. How does Vaisesika explain the formation of more complex structures out of individual atoms? Vaisesika asserts the presence of two factors. The first is a physical property residing in atoms themselves that propels two of the same type to form binaries, three binaries to form a triad, and so on. In other words, atoms possess an innate propensity to aggregate. This idea is thus a forerunner of the modern concept of van der Waal forces. The second factor involved is an unseen principle (*adrishta*) – a vague and somewhat metaphysical concept that seems totally out of place with the rationalist teachings of Vaisesika. It is most likely that this did not form part of the original doctrine and is a later insertion. Actually, very little of Kanada's original work survives; what we have today is attributions, together with commentaries from later members of the Vaisesika tradition, most notably from Prasastapada (6th century CE). There is another angle to this phenomenon of retreat from rationalism. Recall that Indian cosmography is one of periodic dissolution (*pralaya*) and reconstitution taking place indefinitely. Vaisesika doctrine adds a gloss to this picture. At pralaya, the bond between atoms is broken. We now have a collection of isolated, individual atoms; as such the world is beyond perception. In this state, the natural tendency of atoms to congregate is believed to be suspended. What then causes the reinstatement of the atomic tendency to congregate and form more complex structures? Here the unseen principle comes in handy. As one can guess, this principle would eventually be elevated to the status of a divine will. Indeed, Udayana (950 CE) constructed 'a proof of the existence of God' on this basis. If nothing else, this situation mirrors the sad decline of rationalism together with the concomitant rise of theism in the Indian intellectual tradition.

Atomic doctrine also appears in the *Naya sutra* of Aksapada Gautama, although this work is more concerned with logic[14]. The Buddhist world-view also involves atomism. Atoms are of four types, namely those of air, water, fire and earth, as in *Vaisesika*. However, a point of difference between the Buddhist and the *Vaisesika* doctrines ought to be noted[10]. In the Buddhist view a composite body is looked upon as a mere aggregate of its constituent atoms (note 16)[7]. In contrast, Vaisesika (also the atomism of the Jains) recognize the existence of properties of the composite body that result from the manner in which the atoms are put together and organized. We can easily see the difference between these two views by considering the case of (liquid) water versus (solid) ice. Both are made of water atoms, but the bulk material has, quite obviously, distinct properties. Thus Vaisesika is a more developed scheme. This also strengthens the case that Vaisesika came after the Buddha.

Although our understanding, as to what the ultimate constituents of matter are, has evolved and changed in the course of the last two millennia, the idea itself has proved to be resilient and correct. In this sense, the atomic hypothesis is perhaps the most remarkable idea of premodern times. The ancients did not have the ability to perform controlled experiments; nor did they possess the mathematical tools adequate to address this issue. Despite these limitations, through a combination of imagination and reasoning, they stumbled on the idea of atoms – an idea whose validity has been established in the course of the last century and more.

Where do we stand today? The ultimate constituents of matter are believed to be the following: the six leptons, six (triplets) quarks and particles that transmit forces between them. The latter are four carriers of electroweak forces, the eight gluons that carry strong forces between quarks and finally, the Higgs boson that is needed to implement the theoretical scheme of broken gauge symmetry. It is not entirely clear if graviton, the hypothetical particle that could carry gravitational forces, should or should not be included in the list of elementary entities. This is because it is not certain that the gravitational field, described by the (nonlinear) Einstein field equations, can be quantized, in a mathematically consistent way. In any case, our ultimate entities do not all possess the properties that the ancients attributed to their atoms. Thus, not all of them are indestructible; only some are, the rest undergo spontaneous decay. What about visibility? We should first reinterpret the word to mean not just the ability to detect with our sense organs, but more generally to observe via modern scientific instruments. Then only quarks and gluons are invisible (note 17), the rest are visible.

## Notes

1. Heraclitus is more famous for his dialectical logic: All things are in perpetual flux; everything comes into being, yet forthwith ceases to be. This doctrine is strikingly similar to the one Sunyata developed later by the Buddhist Nagarjuna (200 CE).
2. Almost two and a half millennia later, Heraclitean dialectics would be revived by Hegel and then by Marx and his followers to develop the concept of dialectic materialism.
3. *In Chuang Tzu*, composed in the 4th century BCE by Chuang Chao, it is suggested that water is the primary element.
4. See the article by W.C. Archer in ref. 5.
5. In the British Museum. Reproduced in the book by Bor and Reschke[6].
6. Translated from Pali by T.W. Rhys Davids under the heading *Dialogues of the Buddha* (three volumes) (ref. 8).
7. Since Ajatsatru ascended the throne at Magadha (after committing patricide), in 493 BCE (ref. 9) and died in 461 BCE, whereas the Buddha died in 486 BCE, Pakuda must have flourished during the period 493–486 BCE.
8. Mahavira means the great hero, so named because he struggled against and triumphed over his own inhibitions and temptations, needed to become free to lead the life of an ascetic renunciate.
9. The smallest unit of life.
10. A strikingly similar concept was developed by the Greek philosopher Anaxagorous (born 500 BCE). Mind (*nous*) is a substance that enters into the composition of living things. It is the source of all motion. One point of difference should be noted here. Anaxagorous believed that matter is infinitely divisible, i.e. non-atomic.
11. It might be instructive to revisit the six heterodox preachers mentioned in the *Digha Nikaya*. King Ajatsatru of Rajagriha asked the same question to each of them: Is there an immediate fruit, visible in this very world, of the life of a recluse? None of the six answered the question directly, but propounded their respective philosophy instead. It would thus appear that the common unsaid answer was simply that the reward was in the knowledge they had discovered.
12. This is apparently connected with the ethical principle of free will (as against determinism).
13. A beautiful poetic description of the ten stages of this evolution is presented in Jayadeva's *Gita Govindam*. Here the 'spirit of the Lord' is manifested first as a fish, next as a turtle, then as a wild boar and so on till the human stage is reached (ref. 13).
14. Indian revolutionary and comintern (communist international) functionary, who grew disillusioned and severed connection with the communist movement. Founded the Radical Humanist Movement in India. However, he remained a Marxist.
15. Vaisesika is derived from visesa meaning particular, distinctive.
16. The connection between a composite body and its constituents is also discussed in the work Milindapanha, which summarizes the conversations between King Menander and the Buddhist monk Nagasena. Menander was the Greek king of Punjab. According to Nagasena, a body is no more than the sum total of its parts (ref. 7).
17 It is believed (but not proved) that quarks and gluons are permanently confined inside hadrons. In other words, asymptotic states of single particles for quarks or gluons do not exist.

ACKNOWLEDGEMENTS. The idea of this essay arose out of a conversation with Prof. P. P. Divakaran to whom I am much obliged for his continued interest in this project. I also thank Prof. W. D. McGlinn for many discussions Prof. Christopher Nadon Claremont McCenna College for a tutorial on Lucretius, Manjit and Kiran Bhatia for pointing out the presence of the concept of 'Chaurasi Lakh Janam' in the Sikh scriptural literature, and Prof. Don Howard for advice.

S. K. Bose is in the Department of Physics, University of Notre Dame, Notre Dame, Indiana 46556, USA
*e-mail: bose.1@nd.edu